\documentclass[%
 prt,
 reprint,
 superscriptaddress,
 preprintnumbers,
 nofootinbib,
 amsmath,amssymb,
 aps,
]{revtex4-1}

\usepackage{graphicx}
\usepackage{dcolumn}
\usepackage{color}
\usepackage[colorlinks=true,citecolor=blue,urlcolor=blue]{hyperref}
\usepackage{bm}
\usepackage{todonotes}

\begin{document}

\title{Fresh Look at Neutrino Self-Interactions With the Lyman-$\alpha$ Forest: Constraints from EFT and PRIYA Simulations}

\author{Adam He}
\affiliation{%
 Department of Physics and Astronomy, University of Southern California, Los Angeles, CA 90089, USA}
\author{Mikhail M. Ivanov}
\affiliation{Center for Theoretical Physics, Massachusetts Institute of Technology, Cambridge, MA 02139, USA}
\affiliation{The NSF AI Institute for Artificial Intelligence and Fundamental Interactions, Cambridge, MA 02139, USA}
\author{Simeon Bird}
\affiliation{Department of Physics \& Astronomy, University of California, Riverside, Riverside, CA 92521, USA}
\author{Rui An}
\affiliation{%
 Department of Physics and Astronomy, University of Southern California, Los Angeles, CA 90089, USA}
\author{Vera Gluscevic}
\affiliation{%
 Department of Physics and Astronomy, University of Southern California, Los Angeles, CA 90089, USA}

\begin{abstract}

We present the first search for evidence of neutrino self-interaction with two new, state-of-the-art likelihoods for eBOSS Lyman-$\alpha$ data. These are an effective field theory (EFT) likelihood with priors from the Sherwood simulation suite, and a compressed likelihood derived from an emulator built using the PRIYA simulation suite. Previous analyses that combined \textit{Planck} measurements with eBOSS Lyman-$\alpha$ likelihoods based on earlier simulations found a preference for neutrino self-interactions. In contrast, using either of the new eBOSS Lyman-$\alpha$ likelihoods, we find that a joint analysis with the cosmic microwave background (CMB) data from \textit{Planck} prefers a negligible level of neutrino self-interaction, and derive new constraints on the neutrino self-coupling: $\mathrm{log}_{10}(G_\mathrm{eff} \ \mathrm{MeV}^2)=-5.57_{-0.58}^{+0.98}$ for \textit{Planck} + EFT Lyman-$\alpha$, and $\mathrm{log}_{10}(G_\mathrm{eff} \ \mathrm{MeV}^2)=-5.16_{-0.67}^{+1.12}$ for \textit{Planck} + PRIYA Lyman-$\alpha$, at 68\% confidence.
We also consider \textit{Planck} in combination with DESI BAO data, and find that the latter does not provide significant constraining power for neutrino self-interactions. 
\end{abstract}

\preprint{MIT-CTP/5854}
\maketitle

\section{\label{sec:level1}Introduction\protect}

The neutrino sector remains an elusive component of the Standard Model (SM), leaving open the possibility of neutrino interactions with beyond-SM particles. Ground-based experiments that probe neutrino interactions are underway \cite{Aguilar_Arevalo_2021, icecubecollaboration2023measurement, novacollaboration2023measurement, Adamson_2020}. At the same time,  observations from the cosmic microwave background (CMB) and large-scale structure (LSS) provide a powerful, complementary tool to probe couplings between neutrinos and other particles \cite{Choi_2018, Song_2018, Lorenz_2019, Barenboim_2019, Forastieri_2019, Smirnov_2019, Escudero_2020, Ghosh_2020, Funcke_2020, Sakstein_2020, Mazumdar_2020, Blinov_2020, de_Gouv_a_2020, Froustey_2020, Babu_2020, Deppisch_2020, Kelly_2020, Abenza_2020, He_2020, Ding_2020, Berbig_2020, Gogoi_2021, Barenboim_2021, Das_2021, Mazumdar_2022, Brinckmann_2021, Kelly_2021, Esteban_2021, Arias_Arag_n_2021, Du_2021, Carrillo_Gonz_lez_2021, Huang_2021, Sung_2021, Escudero_2021, Choudhury_2021, Roy_Choudhury_2022, Carpio_2023, Orlofsky_2021, Esteban_2021_2, Venzor_2022, Venzor_2023}. 
In particular, Refs. \cite{Kreisch_2020,Kreisch_2022,camarena2023twomode,He_2024} focused on a model in which the onset of neutrino free-streaming is delayed due to self-scattering through a massive scalar mediator $\phi$. 
In this model, the self-interaction rate is parametrized as $\Gamma_{\nu} \propto G_\mathrm{eff}^2T_{\nu}^5$, where $G_\mathrm{eff}$ is a Fermi-like coupling constant describing a four-fermion interaction, and $T_{\nu}$ is the temperature of the neutrino bath.
The delay in neutrino free-streaming leaves a distinct signature on matter clustering in the early and late universe, reducing the size of the sound horizon at recombination, and leading to a higher preferred value of $H_0$ in cosmological analyses \cite{Knox:2019rjx}. Neutrino self-scattering also alters the amount of phase shift in the CMB acoustic peaks caused by neutrino free-streaming, as compared to the standard cosmological model \cite{Bashinsky_2004,Baumann_2016,Follin_2015,montefalcone2025freestreamingneutrinosphaseshift,Kreisch_2020}. Lastly, the self-interactions produce a scale-dependent suppression of the linear matter power spectrum at small scales (Fig. \ref{fig:power spectra}), especially relevant for LSS tracers of high values of $k$.

Previous studies found that a sizable value of the self-coupling constant $G_\mathrm{eff}\sim(10 \ \rm{MeV})^{-2}$ is compatible with the CMB anisotropy data at the $2-3\sigma$ level \citep{Cyr_Racine_2014, Archidiacono_2014, Lancaster_2017, Oldengott_2017,Kreisch_2020,Kreisch_2022}. This model is thus characterized by a bimodal distribution in probability space, with one mode centered at negligible values of $G_\mathrm{eff}$ (dubbed ``moderately-interacting mode'' MI$\nu$, consistent with the standard cosmology), and another mode centered at $G_\mathrm{eff}\sim 0.03\ \mathrm{MeV}^{-2}$ (dubbed ``strongly-interacting mode'' SI$\nu$). The large coupling strengths characteristic of the SI$\nu$ mode were found to simultaneously alleviate the Hubble tension and the $S_8$ tension, the latter representing a disagreement between CMB and LSS measurements of the late-time amplitude of matter fluctuations quantified by the $S_8$ parameter \cite{Abdalla_2022,Di_Valentino_2021_S, Amon_2022,Preston:2023uup, Poulin_2022,rogers2023ultralight,he2023s8}.

In a recent analysis of the Lyman-$\alpha$ data using a compressed likelihood derived from the extended Baryonic Oscillation Sky Survey (eBOSS) \cite{Chabanier_2019_lya, goldstein2023canonical}, 
the DES $S_8$ prior, and the 
EFT-based full-shape power spectrum and bispectrum BOSS likelihood~\cite{Ivanov_2020,Philcox:2021kcw},
we found a strong ($>5\sigma$) preference for a delay in neutrino free-streaming \cite{He_2024}. The compressed Lyman-$\alpha$ likelihood \cite{Chabanier_2019_lya, goldstein2023canonical} was derived from a set of hydrodynamical simulations described in \cite{Palanque_Delabrouille_2015, Borde_2014}. 

Recently, however, an entirely new likelihood was developed based on a new simulation suite, PRIYA \cite{Bird_2023, Fernandez_2024}. When applied to the same eBOSS data, this likelihood prefers a lower amplitude of matter fluctuations, $S_8$ and a higher value of the power spectrum slope, $n_s$ (see also similar results from a recent analysis by the eBOSS collaboration \cite{walther2024emulatinglymanalphaforest1d}). As discussed in Ref.~\cite{Fernandez_2024}, this shift is likely due to a combination of factors, including relatively inaccurate emulation based on a polynomial expansion around a best-fit simulation, an internal tension between different redshift bins ($2.2 \leq z < 2.6$ and $2.6 \leq z \leq 4.6$), and the constraints on simulations imposed by computational limits a decade ago.

In addition, a second likelihood is now available, based on the effective field theory (EFT) of large scale structure (LSS) \cite{Ivanov:2023yla}, based on the 
perturbative  
symmetry-based 
three-dimensional modeling of the 
Lyman-$\alpha$ forest flux field and 
non-linear bias priors calibrated with the Sherwood simulation \cite{Bolton_2016,Givans_2022}. This likelihood, using a reduced set of Lyman-$\alpha$ data in a restricted redshift range ($3.2 \leq z \leq 4.2$), produces a similar set of best-fit cosmological parameters~\cite{Ivanov:2024jtl}.

Motivated by availability of the new likelihoods, we re-examine the evidence for neutrino self-interactions. We include results from both a compressed likelihood derived from the PRIYA-based emulator model \cite{Fernandez_2024} and the full-shape likelihood using EFT with priors from Sherwood~\cite{Ivanov:2023yla}. Both likelihoods are applied in combination with CMB power spectrum measurements from \textit{Planck} \cite{Planck2018_V}. We finally consider the newly-released baryon acoustic oscillation (BAO) data from the Dark Energy Spectroscopic Instrument (DESI) \cite{desicollaboration2024desi2024vicosmological}, as an additional geometric measure of the sound horizon that has sensitivity to neutrino free-streaming delay. 

We find that the compressed Lyman-$\alpha$ likelihood derived from PRIYA, when analyzed in tandem with \textit{Planck} CMB measurements, prefers the standard cosmological model over a neutrino self-scattering model, although strong neutrino self-coupling remains compatible with the two datasets at the $2\sigma$ level. The EFT-based Lyman-$\alpha$ likelihood analyzed in tandem with \textit{Planck} CMB measurements disfavors strong neutrino self-coupling, leading to an upper limit on the interaction strength $G_{\rm{eff}}<10^{-4.4} \ \rm{MeV}^{-2}$, at 95\% confidence.
Adding DESI BAO data to \textit{Planck} does not significantly alter the constraints available from \textit{Planck} alone. The key result is shown in Fig.~\ref{fig:Geff_1D}.

This paper is organized as follows. In Sec.~\ref{sec:formalism}, we describe the neutrino self-scattering cosmology and the EFT of LSS in the context of neutrino self-interactions. 
We describe the likelihoods and datasets used in Sec.~\ref{sec:data} and the analysis in Sec.~\ref{sec:analysis}. We present our results in Sec.~\ref{sec:results}, and discuss their implications and conclude in Sec.~\ref{sec:discussion}.


\section{\label{sec:methods}Method\protect}

\subsection{\label{sec:formalism}Neutrino Self-Interaction Phenomenology\protect}

\begin{figure}[!t]
\includegraphics[scale=0.54]{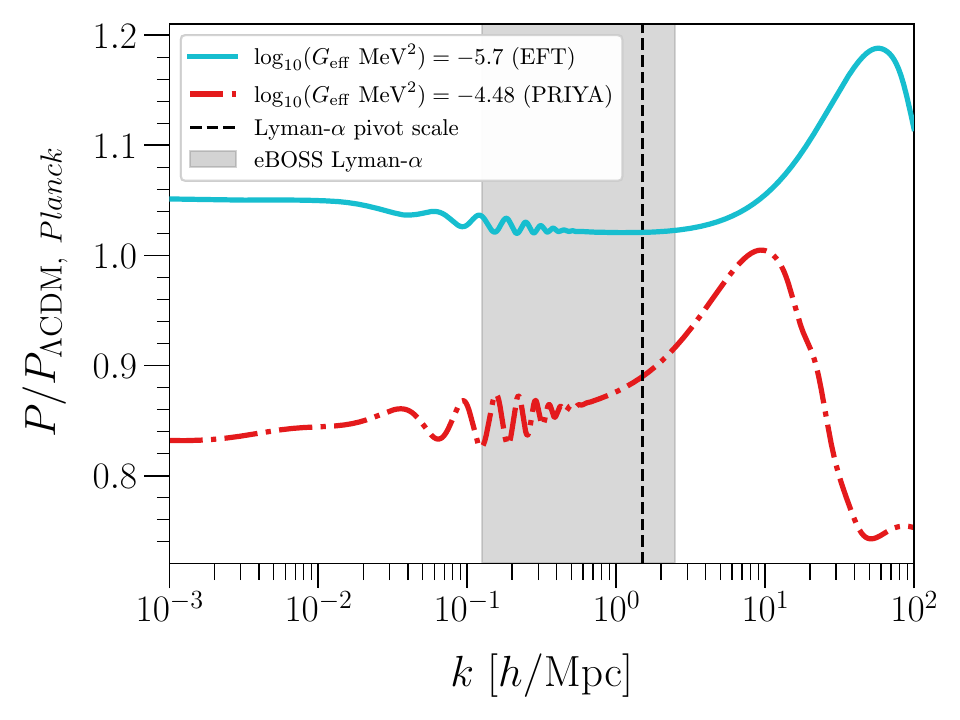}
\caption{\label{fig:power spectra}Ratio of the linear matter power spectra for a self-interacting neutrino cosmology and $\Lambda$CDM cosmology, using best-fit parameters (listed in Tables~\ref{tab:EFT constraints} and~\ref{tab:priya constraints MIv}) from an analysis of Planck + PRIYA Lyman-$\alpha$ data (red) and Planck + EFT Lyman-$\alpha$ data (teal). The gray shaded region indicates the approximate range of scales probed by the eBOSS Lyman-$\alpha$ forest, while the dashed vertical line corresponds to the wavenumber at which the compressed PRIYA likelihood is evaluated.}
\end{figure}

Neutrino self-interactions may be parameterized as $\Gamma_\nu \propto G_\mathrm{eff}^2 T_\nu^5$, where $\Gamma_\nu$ is the neutrino self-interaction rate, $G_\mathrm{eff}$ is a coupling constant describing a four-fermion interaction, and $T_\nu$ is the temperature of the neutrino bath. These interactions alter the Boltzmann equations for the neutrino multipoles; see Ref.~\cite{Kreisch_2020} for a detailed review. To solve these  Boltzmann equations in the presence of neutrino self-interactions, we modify the Boltzmann solver \texttt{CLASS} \cite{Blas_2011} and parametrize by the self-interaction strength $G_\mathrm{eff}$, as in \cite{He_2024}. The effects of the neutrino self-interactions on the linear matter power spectrum, as well as the sensitivity range of the Lyman-$\alpha$ data in $k$ space, are illustrated in Fig.~\ref{fig:power spectra}.

We implement this modeling pipeline into a modified version of \texttt{CLASS-PT}~\citep{Chudaykin2020}\footnote{\url{https://github.com/Michalychforever/CLASS-PT}}, which calculates nonlinear, 1-loop corrections to the linear power spectrum and outputs the 1D Lyman-$\alpha$ flux power spectrum (FPS). \texttt{CLASS-PT} is based on the effective field theory (EFT) of large-scale structure, which is itself based on robust physical principles like symmetries and dimensional analysis. Tests of EFT included, e.g. the field-level comparison with the Sherwood simulations \cite{debelsunce2025modelingcosmologicallymanalphaforest}. \texttt{CLASS-PT} for Lyman-$\alpha$ forest power spectrum predictions has also been tested on the high-fidelity Sherwood \cite{Ivanov:2024jtl} and ACCEL2 \cite{deBelsunce:2024rvv} simulations. Our EFT pipeline was shown to reproduce the input values of the cosmological parameters $\sigma_8$ and $n_s$ from the LACE 1D emulator power spectrum data, which closely matches the eBOSS dataset \cite{Ivanov:2024jtl}.

\texttt{CLASS-PT} employs EFT to model the FPS up to scales $k \sim 10 \ h\mathrm{Mpc}^{-1}$, at redshifts spanning $3.2 < z < 4.2$. eBOSS 2014 measures the FPS at redshifts $2.2 < z < 4.6$; we omit low redshifts $2.2 < z < 3.2$ and high redshifts $4.2 < z < 4.6$, as the scales probed by the forest are redshift-dependent. Lower redshifts probe smaller scales where structure formation is more non-linear, and thus more bias parameters are required for an accurate model. The Sherwood priors for the EFT likelihood
are thereby not sufficient to accurately describe the eBOSS data in the omitted redshift ranges \cite{Ivanov:2024jtl}. In addition, at even lower redshifts $z\lesssim 2.5$, 
EFT is formally 
applicable only on scales $k\lesssim 1~h$Mpc$^{-1}$, which results in a 
loss of information as compared to 
the 
higher redshifts. In principle, the simulation priors used by EFT should be altered to account for neutrino self-scattering. However, the distribution of matter in a neutrino self-interaction scenario evolves just as in $\Lambda$CDM, but with an altered initial power spectrum (see Fig.~\ref{fig:power spectra}), as neutrino self-interactions only affect matter perturbations prior to recombination. We thus use the standard implementation of \texttt{CLASS-PT} to calculate late-time observables like the FPS in the presence of neutrino self-interaction.

The PRIYA Lyman-$\alpha$ likelihood we employ in this work (see Sec. \ref{sec:data}) spans redshifts $2.6 < z < 4.6$ and incorporates a prior on the thermal history of the intergalactic medium gas\footnote{We discard the redshift bins spanning from $2.2 \leq z < 2.6$ due to an internal tension between this low redshift bin and the higher redshift bins within eBOSS data; see Ref.~\cite{Fernandez_2024} for a detailed discussion.}. This compressed likelihood is derived from the PRIYA-simulation based emulator discussed in \cite{Fernandez_2024}. This inference framework has been tested on simulated data, and it successfully recovers the known input parameters \cite{Fernandez_2024}. We fit a 2D Gaussian distribution to posterior constraints on the  amplitude and slope of the linear matter power spectrum at $z_P=3$ and $k_P = 0.009$ s/km $\approx 1 \ \mathrm{Mpc}^{-1}$ from the eBOSS $+ T_0$ chain in \cite{Fernandez_2024}.

Note that the simulations assume $\Lambda$CDM, and thus do not explicitly incorporate the effects of neutrino self-interactions. However, as shown in Ref.~\cite{Pedersen_2023} \cite[see also][]{goldstein2023canonical,rogers20245}, we find a $\Lambda$CDM cosmology with rescaled parameters that can accurately mimic the best-fit power spectrum from the neutrino self-scattering model, and find negligible difference between the two spectra (Appendix \ref{Appendix:rescale}). Thus the use of the compressed likelihood in the neutrino self-scattering scenario is justified for our purpose.

\subsection{\label{sec:data}Data\protect}

We use the following data sets in this study:
\begin{itemize}
    \item \textbf{\textit{Planck}}: \textit{Planck} 2018 \texttt{plik\_lite} high-$\ell$ TT/TE/EE likelihood, with the \texttt{commander} low-$\ell$ TT likelihood and the \texttt{SimAll} low-$\ell$ EE likelihood \citep{Planck2018_V}.
    \item \textbf{Lyman-$\alpha$}: 1D Lyman-$\alpha$ flux power spectrum from SDSS DR14 BOSS and eBOSS quasars.
    \begin{itemize}
    \item \textbf{EFT}:
    An EFT-based full-shape 
    likelihood presented in~\cite{Ivanov:2024jtl} (based on \cite{Ivanov:2023yla}), spanning redshifts $3.2 < z < 4.2$.
    \item \textbf{PRIYA}:
    A compressed likelihood presented as a Gaussian 
    prior on the 
    model-independent amplitude and slope of the power spectrum at a pivot redshift $z_p=3$ and wavenumber $k_p=0.009 \ \mathrm{s/km} \sim 1 \ \mathrm{Mpc}^{-1}$~\citep{Bird_2023}.\footnote{\url{https://github.com/sbird/lya_emulator/blob/master/lyaemu/redlikelihood.py}}
    \end{itemize}
    \item \textbf{DESI}: BAO data from the Dark Energy Spectroscopic Instrument year 1 data release (DR1)~\cite{desicollaboration2024desi2024vicosmological}, at redshift $z>0.8$. Specifically, we
    use the custom likelihood described in~\cite{Chen:2024vuf}. We do not use the DESI LRG BAO DR1 data in order to make 
    our analysis more conservative 
    in light of the discrepancy
    between LRG BAO DR1
   and eBOSS LRG BAO measurements~\cite{eBOSS:2020yzd},
   see e.g.~\cite{Chudaykin:2024gol}.
    \item \textbf{lens}: combined lensing power spectra from Data Release 6 of the Atacama Cosmology Telescope (ACT) and \textit{Planck} PR4 \texttt{NPIPE}  \citep{Madhavacheril_2024,Qu_2024,Carron_2022}. This data is  included in any analysis involving \textit{Planck}, so we do not make note of it when we label our data combinations.
\end{itemize}

\subsection{\label{sec:analysis}Analysis\protect}

In anticipation of the bimodal posterior distribution that is typically found in analyses of this model, we use the \texttt{MultiNest} \cite{Feroz_2008} sampler in tandem with the \texttt{MontePython} \cite{Brinckmann_2018,Audren_2012} sampler to perform likelihood analyses with our modified \texttt{CLASS-PT} code. We assume broad, flat priors on the six standard cosmological parameters $\{\omega_{\mathrm{b}}, \omega_{\mathrm{DM}}, 100\theta_{\mathrm{s}}, \tau_{\mathrm{reio}}, \ln(10^{10}A_{\mathrm{s}}), n_{\mathrm{s}}\}$. For the interacting neutrino cosmology, we follow \cite{He_2024} and vary two more parameters: $\mathrm{log}_{10}(G_{\mathrm{eff}} \ \mathrm{MeV}^{-2})$ and $\sum m_{\nu}$. Analogously, we vary $\sum m_{\nu}$ in our $\Lambda$CDM model and label it $\Lambda$CDM+$\sum m_\nu$. We hold the effective number of relativistic species $N_{\mathrm{eff}}$ fixed at its standard value 3.046\footnote{The standard value of $N_\mathrm{eff}$ is now typically chosen as 3.044. We test minor changes to $N_\mathrm{eff}$ such as this on our model, however, and do not find any noticeable change to our results.}.
We vary the sum of the mass of neutrinos $\sum m_\nu$, and we assume the neutrino sector consists of one massive neutrino containing all of the mass, and two massless neutrinos. 

We list our adopted prior ranges in Appendix~\ref{Appendix:priors}. As in \cite{He_2024}, we limit the lower bound of our prior on $\mathrm{log}_{10}(G_{\mathrm{eff}} \ \mathrm{MeV}^{-2})$ to $-7$. This is because the Lyman-$\alpha$ data probes down to scales $\sim 1 \ \mathrm{Mpc}^{-1}$, which are only sensitive to interaction strengths $\mathrm{log}_{10}(G_{\mathrm{eff}} \ \mathrm{MeV}^{-2}) > -6$. Thus, values of $\mathrm{log}_{10}(G_{\mathrm{eff}} \ \mathrm{MeV}^{-2})$ that are less than $-6$ are effectively equivalent to $G_{\mathrm{eff}}= 0$ in this analysis.

In \texttt{MultiNest}, we fix the number of live points to 1000, the target sampling efficiency to 0.8, and the accuracy threshold for the log Bayesian evidence to 20\%, values that optimize our analysis for parameter estimation. We show the key resulting posterior probability distributions in Figs.~\ref{fig:Geff_1D} and \ref{fig:slope_amplitude}, discussed in detail in the following.

\section{\label{sec:results}Results\protect}

\begin{figure}[!t]
\includegraphics[scale=0.55]{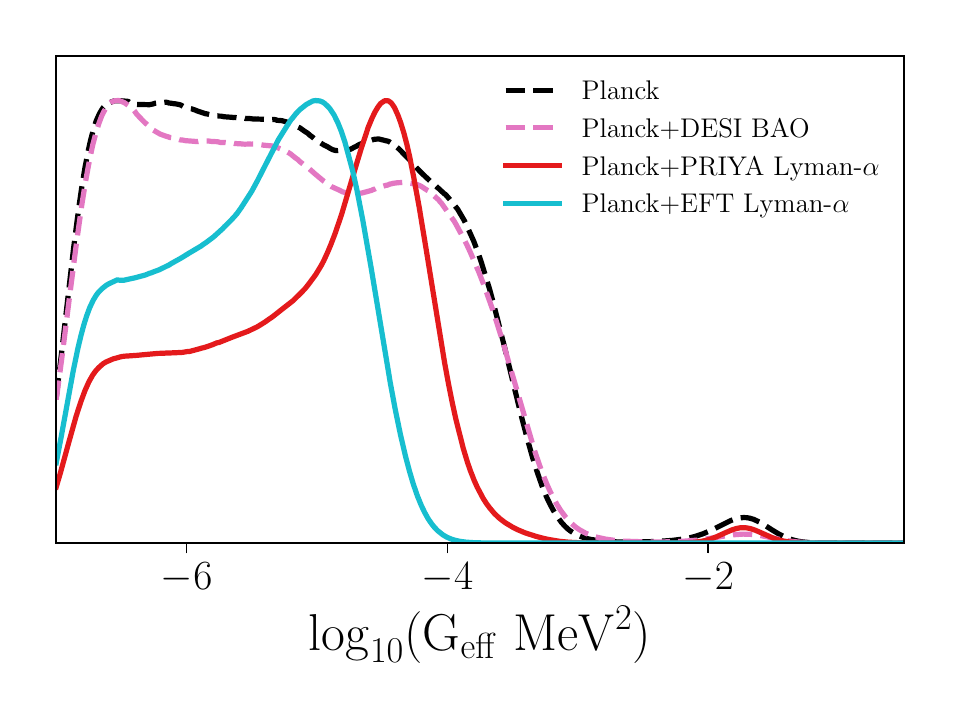}
\caption{\label{fig:Geff_1D}1D posterior probability distributions for the neutrino self-coupling parameter $G_{\rm{eff}}$ from four different analyses: \textit{Planck}--only, \textit{Planck} + PRIYA Lyman-$\alpha$, \textit{Planck} + EFT Lyman-$\alpha$, and \textit{Planck} + DESI BAO. In each analysis, a negligible level of neutrino self-interaction ($G_{\rm{eff}} < 10^{-3} \ \rm{MeV}^{-2}$) is preferred over strong neutrino self-coupling ($G_{\rm{eff}} > 10^{-2} \ \rm{MeV}^{-2}$). }
\end{figure}

In Fig.~\ref{fig:Geff_1D}, we show 1D posterior probability distributions for the neutrino self-coupling parameter $G_{\rm{eff}}$, obtained by analyzing four different combinations of likelihoods: \textit{Planck}--only, \textit{Planck} + PRIYA Lyman-$\alpha$, \textit{Planck} + EFT Lyman-$\alpha$, and \textit{Planck} + DESI.
Consistent with previous work, in the \textit{Planck}--only case we find that the posterior probability distribution is bimodal, with the weak neutrino self-coupling statistically preferred over the strong self-coupling at the $3\sigma$ level, leading to an upper limit of $\rm{log}_{10}(G_{\rm{eff}} \ MeV^2) < -3.48$, at 95\% confidence. 
When we use the PRIYA Lyman-$\alpha$ likelihood in conjunction with \textit{Planck} data, the strong self-coupling is marginally less penalized; overall, the weak self-coupling mode (consistent with a standard cosmology and a negligible delay in neutrino free-streaming) is preferred over the strong self-coupling mode at the $2\sigma$ level, with $\Delta\chi^2=\Delta\chi_{\rm{SI}\nu}^2-\Delta\chi^2_{\rm{MI}\nu}=+5.38$ between the two modes\footnote{We acknowledge that the $\chi^2_{\rm{min}}$ value from one minimization run may not return the true global best-fit minimum, as discussed in   \cite{escudero2025statusneutrinocosmologystandard}; therefore, our $\Delta\chi^2$ comparison should only be used as an approximate metric.}. For this data combination, we find the best-fit value for the self-coupling constant $\mathrm{log}_{10}(G_\mathrm{eff} \ \mathrm{MeV}^2)=-5.16_{-0.67}^{+1.12}$.
When the EFT Lyman-$\alpha$ likelihood is analyzed in conjunction with \textit{Planck}, we find that the strong self-coupling mode is disfavored at $>5\sigma$, with a 95\% upper limit on the interaction strength $\rm{log}_{10}(G_{\rm{eff}} \ MeV^2) < -4.4$ and the best-fit value $\mathrm{log}_{10}(G_\mathrm{eff} \ \mathrm{MeV}^2)=-5.57_{-0.58}^{+0.98}$. Finally, when DESI BAO data is analyzed in conjunction with \textit{Planck}, the result is qualitatively the same as a \textit{Planck}--only analysis; therefore, DESI BAO measurements of the sound horizon do not yet have the sensitivity to neutrino self-coupling.

We show full posterior distributions from our analyses in Appendix~\ref{Appendix:posteriors}, and $\chi^2$ comparisons with the extended standard cosmology $\Lambda$CDM+$\sum m_\nu$ in Appendix~\ref{Appendix:chi2}. We show the full set of cosmological parameter constraints from the \textit{Planck} + Lyman-$\alpha$ analyses in Appendix~\ref{Appendix:constraints}. Interestingly, we note that the combined \textit{Planck} + PRIYA Lyman-$\alpha$ analysis shows a $\sim3\sigma$ preference for non-zero sum of neutrino masses, as evidenced in the full posteriors; $\sum m_\nu = 0.33\pm0.11$ eV, at 68\% confidence. This preference arises from the low clustering amplitude, and it is not degenerate with neutrino self-scattering; we leave detailed scrutiny of this result to future work \cite{future}. 

\begin{figure}[!ht]
\includegraphics[scale=0.36]{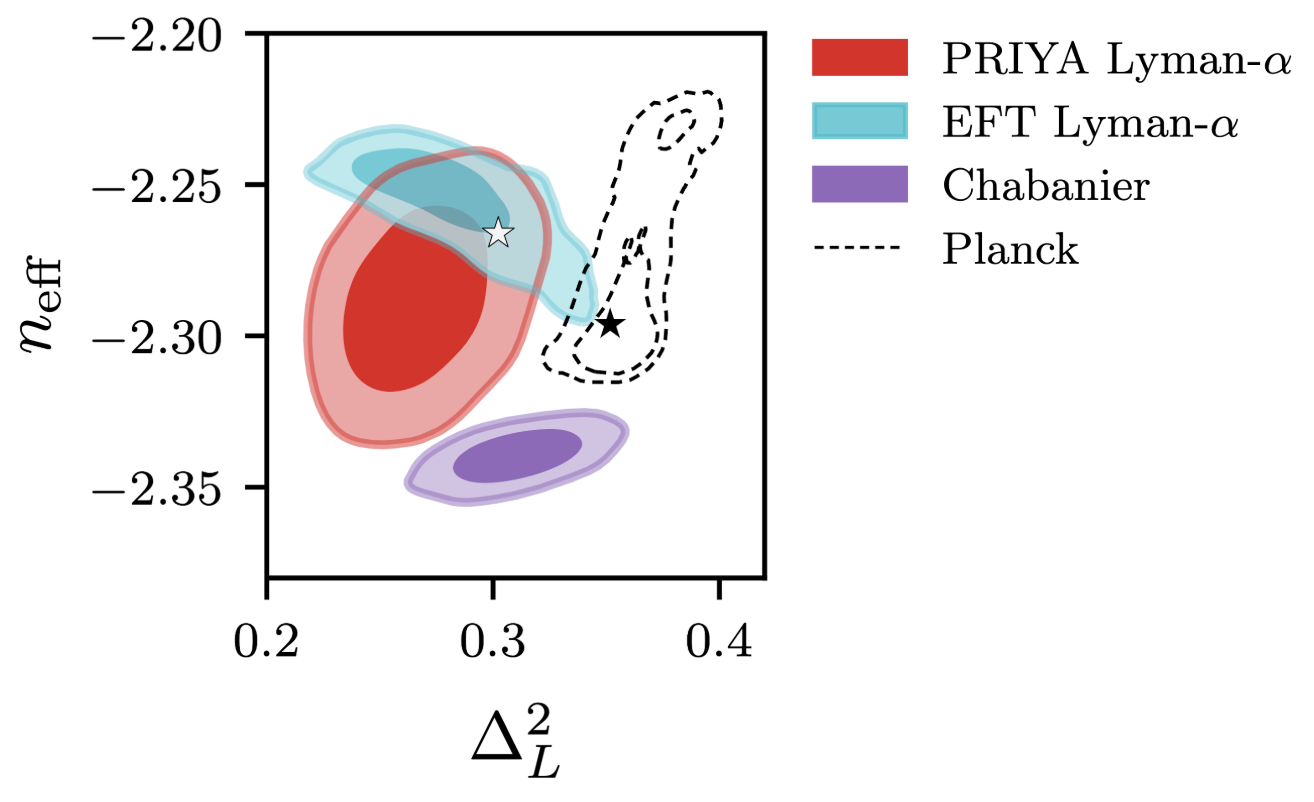}
\caption{\label{fig:slope_amplitude}Slope $n_{\rm{eff}}$ and amplitude $\Delta_L^2$ of the self-interacting neutrino matter power spectrum at $k\sim1 \ \rm{Mpc}^{-1}$, as measured by different data sets. The white star marks the global best-fit from a \textit{Planck}+PRIYA Lyman-$\alpha$ analysis, and the black star marks the best-fit from a \textit{Planck}+EFT Lyman-$\alpha$ analysis, projected to this parameter space. Both Lyman-$\alpha$ likelihoods prefer a shallower power spectrum slope than the Chabanier likelihood, and are therefore less likely to prefer models that tilt the power spectrum at $k\sim1 \ \rm{Mpc}^{-1}$, as in neutrino self-scattering. 
}
\end{figure}

The primary driver for previously-reported preference for strong neutrino self-coupling was the compressed Lyman-$\alpha$ likelihood \cite{Chabanier_2019_lya, goldstein2023canonical} (henceforth referred to as the ``Chabanier likelihood''), calibrated on hydrodynamical simulations described in Refs.~\cite{Palanque_Delabrouille_2015, Borde_2014}, which prefers a steep red slope and low amplitude of the power spectrum at $k \sim 1 \ \rm{Mpc}^{-1}$; see Figs.~\ref{fig:power spectra} and \ref{fig:slope_amplitude} \cite{He_2024,Kreisch_2020,Kreisch_2022}. Strong neutrino self-scattering suppresses $P(k)$ at the relevant scale, allowing the interacting-neutrino model to better fit the Chabanier likelihood, as compared to $\Lambda$CDM. At the same time, the Chabanier likelihood was known to be in tension with \textit{Planck} \cite{rogers20245}; see Fig.~\ref{fig:slope_amplitude}.

In Fig.~\ref{fig:slope_amplitude}, we show 2D posterior probability distributions for the slope $n_{\rm{eff}}$ and amplitude $\Delta_L^2$ of the self-interacting neutrino power spectrum at $z=3$ and $k \sim 1 \ \rm{Mpc}^{-1}$, for different likelihoods. Indeed, the PRIYA Lyman-$\alpha$ likelihood and the EFT-based likelihood are in agreement and both prefer a smaller slope of the power spectrum (corresponding to a higher $n_s$) at $k \sim 1 \ \rm{Mpc}^{-1}$ compared to the Chabanier likelihood; as such, they do not provide support for models that suppress the power spectrum at the pivot scale, such as strong neutrino self-scattering \cite{He_2024}. 
However, the strongly-interacting mode in $G_{\rm{eff}}$ parameter space remains consistent with the PRIYA Lyman-$\alpha$ likelihood (as seen in Fig.~\ref{fig:Geff_1D}), as in this likelihood constraints on the power spectrum slope are weakened due to a degeneracy with parameters of the astrophysical gas model (specifically patchy helium reionization). Thus relatively steep slopes are still marginally consistent with the Lyman-$\alpha$ data. 

The shift we observe within the $n_{\rm{eff}}$---$\Delta_L^2$ space between the Chabanier likelihood and the two new likelihood functions can be understood as follows \cite{Fernandez_2024, walther2024emulatinglymanalphaforest1d}. The simulations used to derive the Chabanier likelihood  \cite{Palanque_Delabrouille_2015,Borde_2014} have significantly smaller box sizes and particle loads than either the PRIYA simulation suite \cite{Bird_2023} or the Sherwood simulations \cite{Bolton_2016}. Second, the Chabanier likelihood relies on an older interpolation technique which fits variation in the flux power spectrum with a polynomial; this technique is significantly less accurate (especially far from the central value) than the interpolation techniques used in the newer simulations, such as Gaussian processes or neural networks.

Moreover, the PRIYA simulations incorporate a patchy helium reionization model that places reionized bubbles with a $\sim 30$ Mpc radius around potential quasars, generating extra power in the FPS on scales that correspond to the size of the bubbles \cite{Bird_2023}. This excess power is degenerate with the slope of the power spectrum at $k \sim 1 \ \rm{Mpc}^{-1}$. Since the parameters of helium reionization remain uncertain, the amplitude of this excess power must be marginalised over in the likelihood, increasing the posterior uncertainty on the matter power spectrum slope $n_\mathrm{eff}$. Alternative patchy helium reionization models could in principle alter the degeneracy between the scale-dependent bias from helium reionization and $n_\mathrm{eff}$, perhaps changing the width of the $n_\mathrm{eff}$ posterior contours. However, Ref.~\cite{Bird_2023} checked that simulation results were insensitive to small model variations, such as varying the size of the bubbles.

Similarly, the EFT-based Lyman-$\alpha$ likelihood is derived from the high-fidelity Sherwood simulations \cite{Bolton_2016}, which have an exceptionally high mass resolution, while the EFT formalism obviates the need for interpolation. 
Fig.~\ref{fig:slope_amplitude}
also shows posterior contours for the slope and amplitude of the power spectrum using the EFT likelihood. As seen in the figure, the EFT-based likelihood prefers shallower (bluer) slopes for $P(k)$ at the pivot scale than the Chabanier likelihood, although it overlaps the posterior constraints from the PRIYA likelihood. The range of slopes preferred by this dataset is in severe tension with those that are characteristic to a strong neutrino self-coupling; this likelihood therefore eliminates the mode corresponding to strong self-interactions in $G_{\rm{eff}}$ parameter space (Fig.~\ref{fig:Geff_1D}).

A comment on the width of the posterior contours of the EFT likelihood is in order. The EFT likelihood uses Lyman-$\alpha$ data only from the redshift range $z \geq 3.2$, yet places much tighter posterior constraints on $n_{\rm{eff}}$ than the PRIYA likelihood, which uses additional data from $2.6 \leq z < 3.2$.
The EFT results should be taken in conjunction with the assumption that the non-linear EFT parameters $b_{\mathcal{O}}$ are fully deterministic cosmology-independent functions of the linear density bias parameter $b_1$.
This assumption is 
inspired by the behavior
of non-linear EFT parameters of 
galaxies and dark matter halos~\cite{Ivanov:2025qie}.
The linear density bias and ``velocity bias'' parameters are varied freely 
in each redshift bin of the 
eBOSS data, 
which allows us to maintain
a sufficient amount of 
flexibility in the fit. 
The relations $b_{\mathcal{O}}(b_1)$ have been calibrated on the Sherwood simulations~\cite{Bolton_2016}. While this calibration is consistent with recent 
precision measurements based on the ACCEL2 hydrodynamical simulation~\cite{Chabanier:2024knr,deBelsunce:2024rvv}, 
we note that neither Sherwood nor ACCEL2 include a realistic model for
patchy helium reionization (although they do both match the evolution of the mean temperature of the intergalactic medium gas) and thus do not model the degeneracy between the heating during helium reionization and the power spectrum slope \cite{Pontzen_2014, Upton_Sanderbeck_2020, Fernandez_2024}. This may account for the difference between the EFT and PRIYA results seen in Fig.~\ref{fig:slope_amplitude}.
It will be interesting to include the
effect of 
patchy helium reionization in the EFT, and more generally, 
to update the EFT likelihood with
simulation-based priors 
calibrated
at the level of the three-dimensional flux field on a set of simulations spanning a large grid of hydrodynamical models, along the lines of the field-level simulation-based priors for galaxy clustering~\cite{Ivanov:2024hgq,Ivanov:2024xgb,Ivanov:2024dgv}. We leave a related study for future work.

\section{\label{sec:discussion}Discussion and Conclusions\protect}

We analyzed the CMB and LSS data in search for neutrino self-scattering that delays the onset of neutrino free-streaming, using two new and updated re-analyses of the Lyman-$\alpha$ data from eBOSS. The two new likelihoods are calibrated on the state-of-the-art simulation suites PRIYA \cite{Bird_2023} and Sherwood \cite{Bolton_2016,Givans_2022}, and use respectively a compressed likelihood based on an emulator \cite{Fernandez_2024} and an EFT-based full-shape likelihood \cite{Ivanov:2024jtl}. 
Previous analyses relied on an earlier compressed likelihood from eBOSS \cite{Chabanier_2019_lya} and found a strong preference for a delay in neutrino free-streaming over the standard cosmology. However, the updated Lyman-$\alpha$ likelihoods both prefer a vanilla $\Lambda$CDM cosmology, providing more stringent constraint on neutrino self-interactions than all previous analyses. Specifically, we find $\mathrm{log}_{10}(G_\mathrm{eff} \ \mathrm{MeV}^2)=-5.57_{-0.58}^{+0.98}$ for \textit{Planck} + EFT Lyman-$\alpha$, and $\mathrm{log}_{10}(G_\mathrm{eff} \ \mathrm{MeV}^2)=-5.16_{-0.67}^{+1.12}$ for \textit{Planck} + PRIYA Lyman-$\alpha$, at 68\% confidence; key results are in Figs.~\ref{fig:Geff_1D} and \ref{fig:slope_amplitude}. 
Furthermore, we find that the BAO data from DESI does not presently have sufficient sensitivity to probe neutrino scattering at a level comparable to CMB and LSS. Finally, we note a preference for a non-vanishing sum of neutrino masses in some datasets, to be explored in detail in a future study.

The significant reduction in preference for neutrino self-interactions we report in this work can be understood as a shift in preferred amplitude and slope of $P(k)$ around $k \sim 1 \ \rm{Mpc}^{-1}$, driven by differences in the simulation suites used to forward-model the Lyman-$\alpha$ forest. Concretely, likelihoods calibrated on previous suites of Lyman-$\alpha$ simulations returned a steep slope and low amplitude of $P(k)$ at the relevant scale, a preference which is removed in newer likelihoods. Consequently, strong neutrino self-interactions, which lead to a low amplitude and steep slope of the power spectrum, are no longer preferred over $\Lambda$CDM, in light of new Lyman-$\alpha$ data.

Although the statistical preference for strong neutrino self-coupling is significantly reduced in this work, we find that strong neutrino self-interactions remain consistent with a combination of CMB and LSS data from the Lyman-$\alpha$ forest. This consistency is largely dependent on this model's suppression of structure at small scales, which will be imminently tested by a multitude of future experiments. For example, upcoming galaxy surveys from the Vera C. Rubin Observatory \cite{VeraRubin2018,Nadler_2019,Nadler_2021} and other observational facilities \cite{DESI:2016fyo,SPHEREx:2014bgr} will tighten current limits on the allowed level of suppression at these scales \cite{Chudaykin:2019ock,Sailer:2021yzm,Nadler_2019, drlicawagner2019}, thereby constraining strong neutrino self-scattering and any other beyond-CDM models that affect the matter power spectrum in this regime. Precise measurements of the CMB anisotropy from the Simons Observatory \citep{SimonsObservatory:2018koc} will also act as a crucial lever arm to constrain the shape of the power spectrum at these scales. 
Finally, this study highlights the importance of the Lyman-$\alpha$ forest in cosmological analyses, especially the high-fidelity likelihoods based on simulation, and incorporating  observational constraints on astrophysical parameters, that we use here. We anticipate that the re-analyses of other beyond-standard cosmologies that relied on the forest measurements from eBOSS will need updates analogous to this study \cite{goldstein2023canonical,rogers20245,He_2024}. 

\section*{Acknowledgements}

VG acknowledges the support from NASA through the Astrophysics Theory Program, Award Number 21-ATP21-0135, from the National Science Foundation (NSF) CAREER Grant No. PHY-2239205, and from the Research Corporation for Science Advancement through the Cottrell Scholars program.
SB acknowledges funding from NASA
ATP 80NSSC22K1897.

\clearpage
\appendix
\onecolumngrid

\section{Rescaled $\Lambda$CDM spectra}
\label{Appendix:rescale}

We verify that the compressed PRIYA Lyman-$\alpha$ likelihood is valid in the context of neutrino self-interactions by plotting a $\Lambda$CDM cosmology that has rescaled parameters which accurately mimic the best-fit power spectrum from our analysis. Because the two spectra are indistinguishable at the 0.5\% level in the range of scales and redshifts probed by the eBOSS Lyman-$\alpha$ data, we may use the reduced likelihood to analyze the self-interacting neutrino model, without running additional simulations.

\begin{figure*}[!htb]
\includegraphics[scale=0.65]{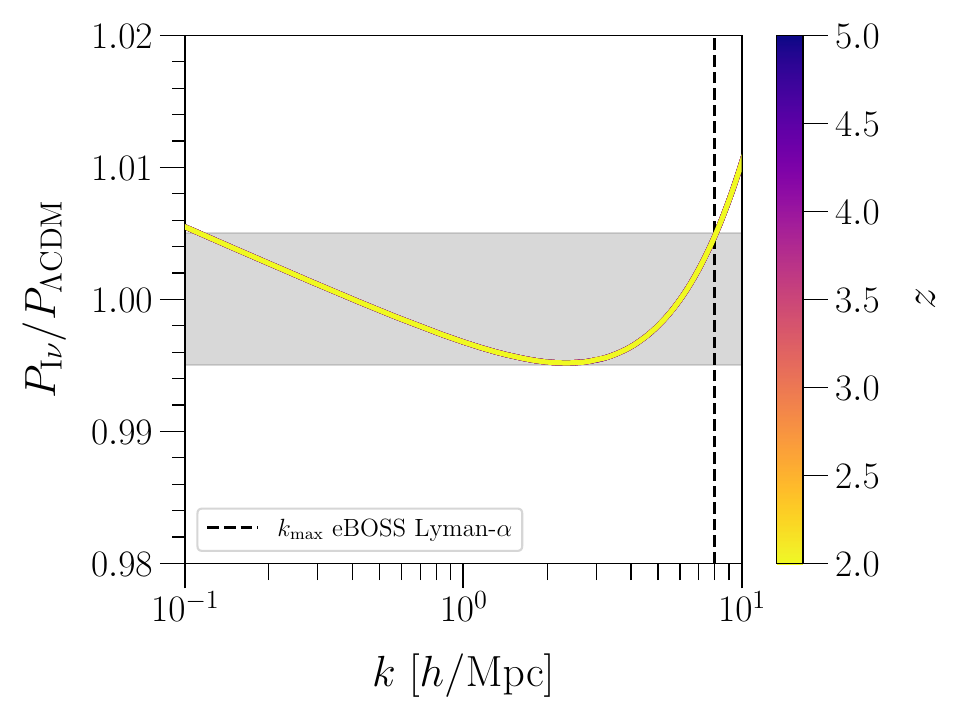}
\caption{\label{fig:rescale} Ratio of the power spectrum for a $\Lambda$CDM cosmology with rescaled parameters and the power spectrum of the best-fit self-interacting neutrino cosmology from a Planck + PRIYA Lyman-$\alpha$ analysis, at redshifts $z=2-5$. The vertical dotted line shows the approximate maximum 3D wavenumber probed by the eBOSS Lyman-$\alpha$ dataset. Both models are indistinguishable at the 0.5\% level in the range of scales and redshifts probed by the eBOSS data, indicated in gray.}
\end{figure*}

\raggedbottom

\pagebreak

\clearpage

\section{Prior ranges}
\label{Appendix:priors}

We list all prior ranges on cosmological parameters in Table~\ref{tab:priors}. Priors for standard cosmological parameters are given in the top half of the table, and priors on \textit{Planck} and EFT bias parameters are given in the bottom half. The superscripts (1), (2), (3), (4), (5), (6) of the bias parameters $b_1,b_{\eta}$ signify the $z=3.2$, $z=3.4$, $z=3.6$, $z=3.8$, $z=4.0$, $z=4.2$ redshift bins, respectively.

Note that our priors extend past some of the parameter ranges used in the PRIYA simulation suite \cite{Bird_2023}. This is because, for our \textit{Planck}+PRIYA analysis, we do not directly use a Gaussian process; rather, we use a Gaussian likelihood fit to the marginalized likelihood contours of the underlying Gaussian process likelihood. Our analysis therefore captures the range of power spectra that fit this Gaussian likelihood, and returns the range of cosmological parameters that create these spectra, which may take on values that are outside the ranges initially used in the simulations.

\begin{table*}[htb!]
\centering
\caption{Adopted prior ranges.} \label{tab:priors}
\begin{tabular}{|c|c|c|c|}
\hline
Parameter & Prior \\
\hline 
\hline
$100~\omega{}_\mathrm{b }$ & $[1.0,3.9]$ \\ [0.5ex] 
$\omega{}_\mathrm{c }$ & $[0.08,0.16]$ \\ [0.5ex] 
$100~\theta{}_\mathrm{s }$ & $[1.03, 1.05]$ \\ [0.5ex] 
$\mathrm{ln}(10^{10}A_\mathrm{s })$ & $[2,4]$ \\ [0.5ex] 
$n_\mathrm{s }$ & $[0.85,1.1]$ \\ [0.5ex] 
$\tau{}_\mathrm{reio }$ & $[0.01,0.25]$ \\ [0.5ex] 
$\mathrm{log}_{10}(G_\mathrm{eff} \ \mathrm{MeV}^2)$ & $[-7,-0.5]$ \\
[0.5ex]
$\sum m_\nu \ [\mathrm{eV}]$ & $[0.0001, 1.5]$ \\ [0.5ex] 
\hline
$A_\mathrm{planck}$ &$[0.9,1.1]$ \\ [0.5ex] 
$b^{(1)}_{1 }$ & $[-1,0]$ \\ [0.5ex] 
$b^{(2)}_{1 }$ & $[-1,0]$ \\ [0.5ex] 
$b^{(3)}_{1 }$ & $[-1,0]$ \\ [0.5ex] 
$b^{(4)}_{1 }$ & $[-1,0]$ \\ [0.5ex] 
$b^{(5)}_{1 }$ & $[-1,0]$ \\ [0.5ex] 
$b^{(6)}_{1 }$ & $[-1,0]$ \\ [0.5ex] 
$b^{(1)}_{\eta}$ & $[-1,1]$ \\ [0.5ex] 
$b^{(2)}_{\eta}$ & $[-1,1]$ \\ [0.5ex] 
$b^{(3)}_{\eta}$ & $[-1,1]$ \\ [0.5ex] 
$b^{(4)}_{\eta}$ & $[-2,0]$ \\ [0.5ex] 
$b^{(5)}_{\eta}$ & $[-2,0]$ \\ [0.5ex] 
$b^{(6)}_{\eta}$ & $[-2,0]$ \\ [0.5ex] 

 \hline
\end{tabular}
\end{table*}
\raggedbottom

\raggedbottom

\pagebreak

\clearpage

\section{Posteriors}
\label{Appendix:posteriors}

\begin{figure*}[!htb]
\includegraphics[scale=0.65]{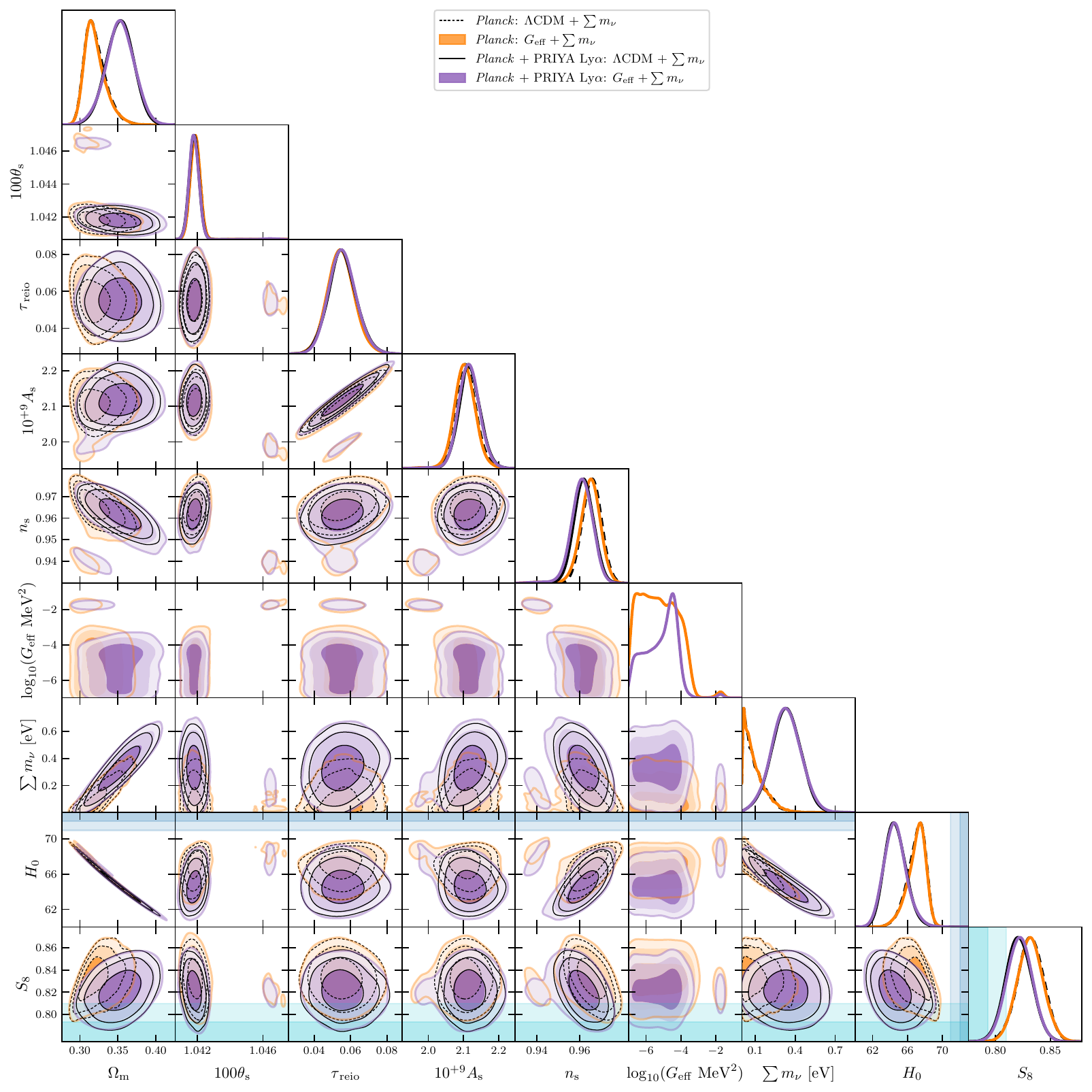}
\caption{\label{fig:lyabird_full_triangle_plot} 68\%, 95\%, and 99\% confidence level marginalized posterior distributions for the $\Lambda$CDM+$\sum m_\nu$ model (black) and the self-interacting neutrino model $G_{\rm{eff}}$+$\sum m_{\nu}$ (colored) from a combined analysis of \textit{Planck}, PRIYA Lyman-$\alpha$, and CMB lensing data and a combined analysis of \textit{Planck} and CMB lensing only. The blue shaded band marks the $\mathrm{SH_0ES}$ measurement of $H_0$, and the teal shaded band marks the DES measurement of $S_8$.}
\end{figure*}

\begin{figure*}[!htb]
\includegraphics[scale=0.65]{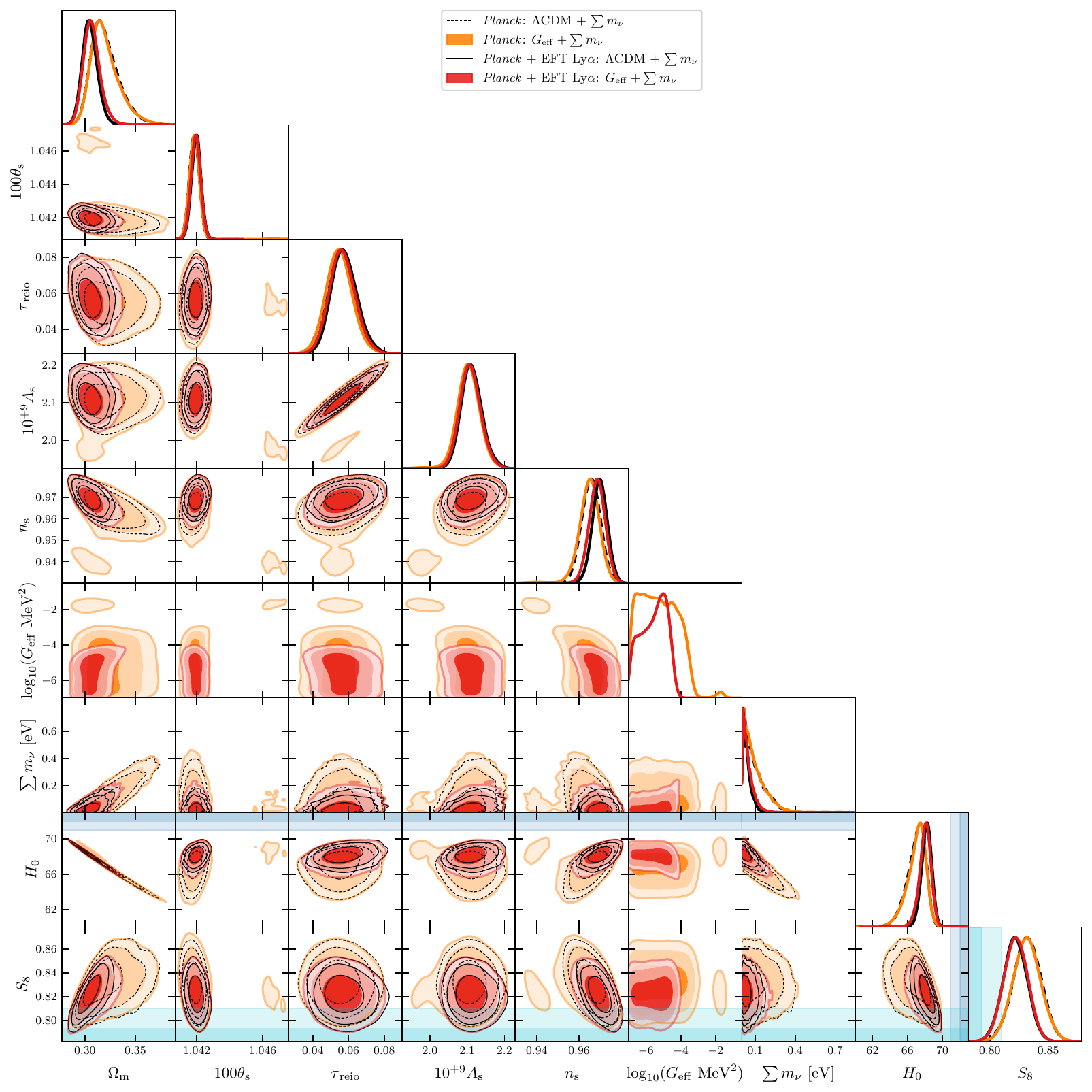}
\caption{\label{fig:eftlya_full_triangle_plot} 68\%, 95\%, and 99\% confidence level marginalized posterior distributions for the $\Lambda$CDM+$\sum m_\nu$ model (black) and the self-interacting neutrino model $G_{\rm{eff}}$+$\sum m_{\nu}$ (colored) from a combined analysis of \textit{Planck}, EFT Lyman-$\alpha$, and CMB lensing data and a combined analysis of \textit{Planck} and CMB lensing only. The blue shaded band marks the $\mathrm{SH_0ES}$ measurement of $H_0$, and the teal shaded band marks the DES measurement of $S_8$.}
\end{figure*}

\begin{figure*}[!htb]
\includegraphics[scale=0.65]{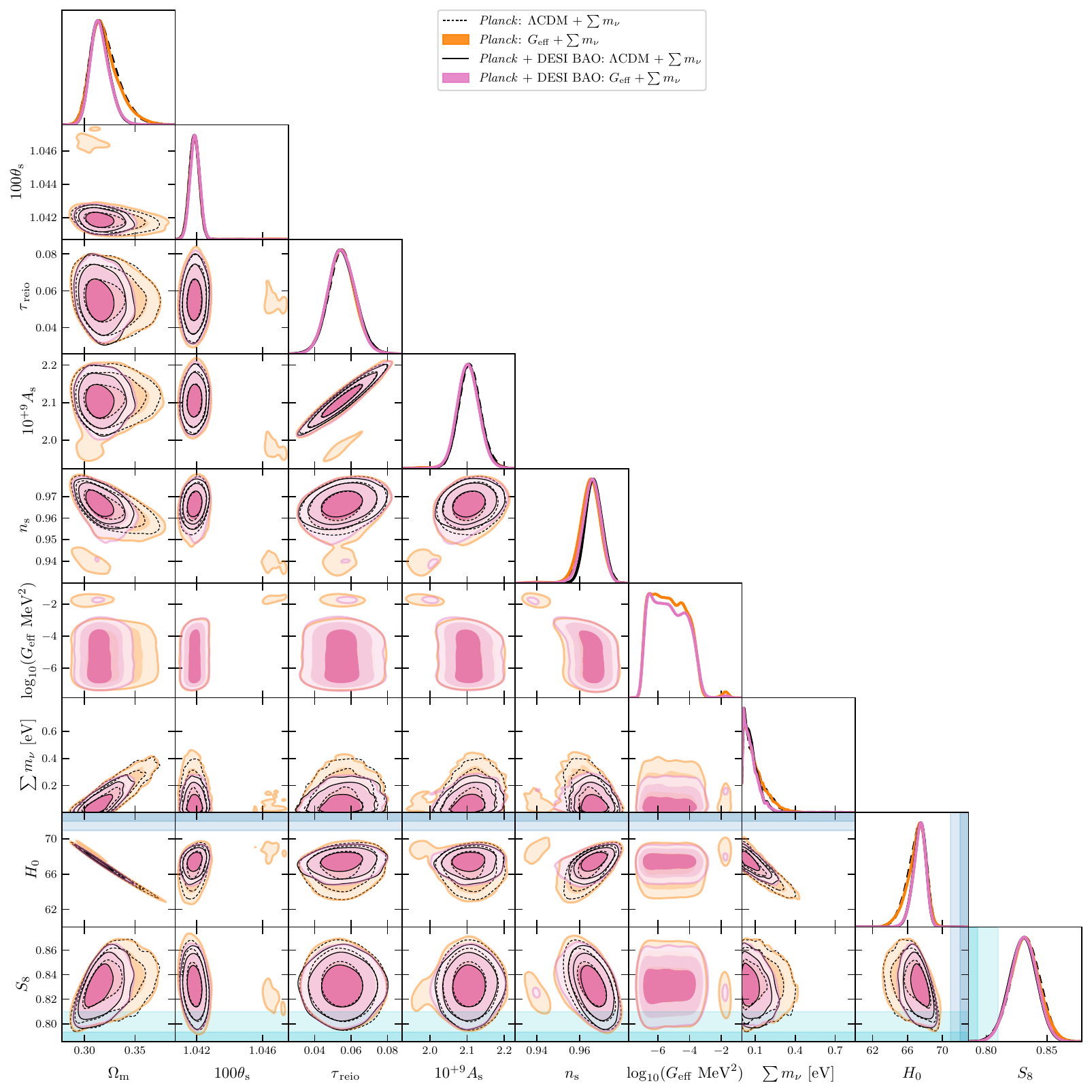}
\caption{\label{fig:desi_full_triangle_plot} 68\%, 95\%, and 99\% confidence level marginalized posterior distributions for the $\Lambda$CDM+$\sum m_\nu$ model (black) and the self-interacting neutrino model $G_{\rm{eff}}$+$\sum m_{\nu}$ (colored) from a combined analysis of \textit{Planck}, DESI BAO, and CMB lensing data and a combined analysis of \textit{Planck} and CMB lensing only. The blue shaded band marks the $\mathrm{SH_0ES}$ measurement of $H_0$, and the teal shaded band marks the DES measurement of $S_8$.}
\end{figure*}

We display full marginalized posterior distributions for all relevant parameters in a \textit{Planck} + PRIYA Lyman-$\alpha$ analysis of the self-interacting neutrino model in Fig.~\ref{fig:lyabird_full_triangle_plot}. We display posteriors for a \textit{Planck} + EFT Lyman-$\alpha$ analysis of the self-interacting neutrino model in Fig.~\ref{fig:eftlya_full_triangle_plot}. We display posteriors for a \textit{Planck} + DESI BAO analysis of the self-interacting neutrino model in Fig.~\ref{fig:desi_full_triangle_plot}.

\raggedbottom

\pagebreak

\raggedbottom

\pagebreak

\raggedbottom

\pagebreak

\clearpage

\section{$\chi^2$ comparisons}
\label{Appendix:chi2}

In Table~\ref{tab:chi2}, we show $\Delta\chi^2$ values between the self-interacting neutrino model $G_{\rm{eff}}$+$\sum m_{\nu}$ and the $\Lambda$CDM+$\sum m_\nu$ model from our analyses. Strong neutrino self-couplings are disfavored by a \textit{Planck} + EFT Lyman--$\alpha$ analysis, at the $>5\sigma$ level.

\begin{table*}[htb!]
\centering
\caption{$\Delta\chi_{\mathrm{min}}^2$ for the self-interacting neutrino model $G_\mathrm{eff}$+$\sum m_\nu$, compared to the $\Lambda$CDM+$\sum m_\nu$ model, under different analyses.} \label{tab:chi2}
\begin{tabular}{|c|c|c|c|c|}
\hline
Data combination & $\Delta\chi_{\rm{MI}\nu}^2$ wrt $\Lambda$CDM+$\sum m_\nu$ & $\Delta\chi_{\rm{SI}\nu}^2$ wrt $\Lambda$CDM+$\sum m_\nu$ \\
\hline 
\hline
\textit{Planck} & $+0.28$ & $+7.49$ \\ [0.5ex] 
\textit{Planck} + PRIYA Lyman--$\alpha$ & $-2.01$ & $+3.37$ \\ [0.5ex] 
\textit{Planck} + EFT Lyman--$\alpha$ & $-0.98$ & $+30.46$ \\ [0.5ex] 
\textit{Planck} + DESI BAO & $-0.21$ & $+10.26$ \\
[0.5ex]
 \hline
\end{tabular}
\end{table*}
\raggedbottom

\section{Cosmological parameter constraints}
\label{Appendix:constraints}

In Table~\ref{tab:EFT constraints}, we display the full set of cosmological parameter constraints from a \textit{Planck} + EFT Lyman-$\alpha$ analysis of the self-interacting neutrino model $G_\mathrm{eff}$+$\sum m_\nu$.
In Table~\ref{tab:priya constraints MIv}, we display cosmological parameter constraints for the moderately-interacting MI$\nu$ mode of the self-interacting neutrino model $G_\mathrm{eff}$+$\sum m_\nu$, from a \textit{Planck} + PRIYA Lyman-$\alpha$ analysis. 
In Table~\ref{tab:priya constraints SIv}, we display cosmological parameter constraints for the strongly-interacting SI$\nu$ mode of the self-interacting neutrino model $G_\mathrm{eff}$+$\sum m_\nu$, from a \textit{Planck} + PRIYA Lyman-$\alpha$ analysis. 

\begin{table*}[htb!]
\centering
\caption{Parameter constraints for a \textit{Planck} + EFT Lyman-$\alpha$ analysis of the self-interacting neutrino model $G_\mathrm{eff}$+$\sum m_\nu$. Constraints for standard cosmological parameters are given in the top half of the table, and constraints on \textit{Planck} and EFT bias parameters are given in the bottom half. The maximum of the full posterior is labeled as ``Best-fit'', and the maxima of the marginalized posteriors are labeled as ``Marginalized max''. The superscripts (1), (2), (3), (4), (5), (6) 
of 
the bias parameters
$b_1,b_{\eta}$
signify the $z=3.2$, $z=3.4$, $z=3.6$, $z=3.8$, $z=4.0$, $z=4.2$ redshift bins, respectively. 
} 
\label{tab:EFT constraints}
\begin{tabular}{|l|c|c|c|c|}
 \hline
Parameter & Best-fit & Marginalized max $\pm \ \sigma$ & 95\% lower & 95\% upper \\ \hline
$100~\omega{}_\mathrm{b }$ &$2.241$ & $2.248\pm0.014$ & $2.22$ & $2.276$ \\
$\omega{}_\mathrm{c }$ &$0.1194$ & $0.1189\pm0.0011$ & $0.1168$ & $0.1211$ \\
$100~\theta{}_\mathrm{s }$ &$1.0418$ & $1.0419\pm{0.00029}$ & $1.0414$ & $1.0425$ \\
$\mathrm{ln}(10^{10}A_\mathrm{s })$ &$3.05$ & $3.049_{-0.014}^{+0.12}$ & $3.023$ & $3.076$ \\
$n_\mathrm{s }$ &$0.967$ & $0.9686\pm0.0038$ & $0.9608$ & $0.9762$ \\
$\tau{}_\mathrm{reio }$ &$0.05806$ & $0.05653_{-0.0076}^{+0.0068}$ & $0.04272$ & $0.07188$ \\
$\mathrm{log}_{10}(G_\mathrm{eff} \ \mathrm{MeV}^2)$ & $-5.699$ & $-5.568_{-0.578}^{+0.981}$ & $-6.908$ & $-4.407$ \\
$\sum m_\nu \ [\mathrm{eV}]$ & $0.026$ & $0.046_{-0.045}^{+0.01}$ & $>0$ & $0.13$ \\
$z_\mathrm{reio }$ &$8.047$ & $7.857_{-0.703}^{+0.712}$ & $6.448$ & $9.305$ \\
$\Omega{}_{\Lambda }$ &$0.692$ & $0.6931_{-0.007}^{+0.0091}$ & $0.6759$ & $0.7091$ \\
$Y_\mathrm{He}$ &$0.2479$ & $0.2479\pm{6e-05}$ & $0.2478$ & $0.248$ \\
$H_0$ &$67.935$ & $68.026_{-0.557}^{+0.71}$ & $66.667$ & $69.295$ \\ 
$10^{+9}A_\mathrm{s }$ &$2.111$ & $2.109_{-0.029}^{+0.026}$ & $2.056$ & $2.168$ \\
$\sigma_8$ &$0.8196$ & $0.813_{-0.0054}^{+0.0097}$ & $0.7947$ & $0.828$ \\
$S_8$ &$0.83$ & $0.822\pm0.01$ & $0.802$ & $0.842$ \\
\hline
$A_\mathrm{planck}$ &$0.99953$ & $1.00127_{-0.00223}^{+0.00227}$ & $0.99688$ & $1.00568$ \\
$b^{(1)}_{1 }$ &$-0.3646$ & 
$-0.3578_{-0.0141}^{+0.0071}$ & $-0.3758$ & $-0.3344$ \\
$b^{(2)}_{1 }$ &$-0.4092$ &
$-0.4039_{-0.0137}^{+0.0079}$ & $-0.4227$ & $-0.3809$ \\
$b^{(3)}_{1 }$ &$-0.4555$ & 
$-0.3352_{-0.1209}^{+0.1088}$ & $-0.4729$ & $-0.2078$ \\
$b^{(4)}_{1 }$ &$-0.5246$ & 
$-0.5194_{-0.0143}^{+0.0102}$ & $-0.5426$ & $-0.4936$ \\
$b^{(5)}_{1 }$ &$-0.6108$ & 
$-0.6009_{-0.0138}^{+0.0111}$ & $-0.624$ & $-0.574$ \\
$b^{(6)}_{1 }$ &$-0.6817$ & 
$-0.6832_{-0.0184}^{+0.0146}$ & $-0.7157$ & $-0.6484$ \\
$b^{(1)}_{\eta}$ &$-0.1162$ & 
$-0.0692_{-0.0677}^{+0.031}$ & $-0.1519$ & $0.0433$ \\
$b^{(2)}_{\eta}$ &$-0.1949$ & $-0.1549_{-0.0684}^{+0.0352}$ & $-0.2433$ & $-0.043$ \\
$b^{(3)}_{\eta}$ & $-0.2663$ & $0.3457_{-0.6397}^{+0.5678}$ & $-0.3409$ & $0.9469$ \\
$b^{(4)}_{\eta}$ &$-0.4415$ & $-0.3944_{-0.0798}^{+0.0584}$ & $-0.5273$ & $-0.2485$ \\
$b^{(5)}_{\eta}$ &$-0.7271$ & $-0.6485_{-0.0842}^{+0.0683}$ & $-0.7969$ & $-0.491$ \\
$b^{(6)}_{\eta}$ &$-0.7894$ & $-0.8098_{-0.1307}^{+0.1017}$ & $-1.0376$ & $-0.565$ \\
\hline
\end{tabular}
\end{table*}
\raggedbottom

\begin{table*}[htb!]
\centering
\caption{Parameter constraints for the moderately-interacting mode, in a \textit{Planck} + PRIYA Lyman-$\alpha$ analysis of the self-interacting neutrino model $G_\mathrm{eff}$+$\sum m_\nu$. The maximum of the full posterior is labeled as ``Best-fit'', and the maxima of the marginalized posteriors are labeled as ``Marginalized max''.
} 
\label{tab:priya constraints MIv}
\begin{tabular}{|l|c|c|c|c|}
 \hline
Parameter & Best-fit & Marginalized max $\pm \ \sigma$ & 95\% lower & 95\% upper \\ \hline
$100~\omega{}_\mathrm{b }$ &$2.218$ & $2.224\pm0.016$ & $2.194$ & $2.256$ \\
$\omega{}_\mathrm{c }$ &$0.1215$ & $0.1212\pm0.0013$ & $0.1186$ & $0.1237$ \\
$100~\theta{}_\mathrm{s }$ &$1.0418$ & $1.0418\pm{0.00029}$ & $1.0412$ & $1.0424$ \\
$\mathrm{ln}(10^{10}A_\mathrm{s })$ &$3.049$ & $3.052^{+0.015}_{-0.013}$ & $3.026$ & $3.081$ \\
$n_\mathrm{s }$ &$0.9604$ & $0.9612\pm0.0048$ & $0.9515$ & $0.9704$ \\
$\tau{}_\mathrm{reio }$ &$0.0534$ & $0.05555_{-0.0079}^{+0.007}$ & $0.04131$ & $0.0712$ \\
$\mathrm{log}_{10}(G_\mathrm{eff} \ \mathrm{MeV}^2)$ & $-4.477$ & $-5.195_{-0.635}^{+1.15}$ & $-6.872$ & $-3.906$ \\
$\sum m_\nu \ [\mathrm{eV}]$ & $0.303$ & $0.331_{-0.107}^{+0.111}$ & $0.11$ & $0.55$ \\
$z_\mathrm{reio }$ &$7.699$ & $7.89_{-0.76}^{+0.753}$ & $6.385$ & $9.433$ \\
$\Omega{}_{\Lambda }$ &$0.6476$ & $0.6468_{-0.0187}^{-0.0182}$ & $0.611$ & $0.6843$ \\
$Y_\mathrm{He}$ &$0.2478$ & $0.2478\pm{6.8e-05}$ & $0.2477$ & $0.2479$ \\
$H_0$ &$64.588$ & $64.576_{-1.339}^{+1.126}$ & $62.161$ & $67.125$ \\ 
$10^{+9}A_\mathrm{s }$ &$2.109$ & $2.117_{-0.031}^{+0.028}$ & $2.061$ & $2.178$ \\
$\sigma_8$ &$0.7624$ & $0.7583_{-0.0197}^{+0.0175}$ & $0.7214$ & $0.7962$ \\
$S_8$ &$0.826$ & $0.822\pm0.012$ & $0.799$ & $0.845$ \\
$A_\mathrm{planck}$ &$1.00215$ & $1.00225\pm{0.0022}$ & $0.99785$ & $1.00666$ \\
\hline
\end{tabular}
\end{table*}
\raggedbottom

\begin{table*}[htb!]
\centering
\caption{Parameter constraints for the strongly-interacting mode, in a \textit{Planck} + PRIYA Lyman-$\alpha$ analysis of the self-interacting neutrino model $G_\mathrm{eff}$+$\sum m_\nu$. The maximum of the full posterior is labeled as ``Best-fit'', and the maxima of the marginalized posteriors are labeled as ``Marginalized max''.
} 
\label{tab:priya constraints SIv}
\begin{tabular}{|l|c|c|c|c|}
 \hline
Parameter & Best-fit & Marginalized max $\pm \ \sigma$ & 95\% lower & 95\% upper \\ \hline
$100~\omega{}_\mathrm{b }$ &$2.264$ & $2.26\pm0.017$ & $2.228$ & $2.293$ \\
$\omega{}_\mathrm{c }$ &$0.1212$ & $0.1208\pm0.0015$ & $0.1178$ & $0.1237$ \\
$100~\theta{}_\mathrm{s }$ &$1.0464$ & $1.0464\pm{0.0004}$ & $1.0456$ & $1.0472$ \\
$\mathrm{ln}(10^{10}A_\mathrm{s })$ &$2.989$ & $2.992_{-0.016}^{+0.015}$ & $2.961$ & $3.025$ \\
$n_\mathrm{s }$ &$0.9383$ & $0.9401_{-0.0051}^{+0.0046}$ & $0.9311$ & $0.9496$ \\
$\tau{}_\mathrm{reio }$ &$0.05449$ & $0.05635_{-0.0086}^{+0.0075}$ & $0.04065$ & $0.07461$ \\
$\mathrm{log}_{10}(G_\mathrm{eff} \ \mathrm{MeV}^2)$ & $-1.707$ & $-1.731_{-0.052}^{+0.068}$ & $-1.866$ & $-1.619$ \\
$\sum m_\nu \ [\mathrm{eV}]$ & $0.206$ & $0.188_{-0.133}^{+0.109}$ & $>0$ & $0.406$ \\
$z_\mathrm{reio }$ &$7.7$ & $7.867_{-0.77}^{+0.791}$ & $6.184$ & $9.567$ \\
$\Omega{}_{\Lambda }$ &$0.6787$ & $0.6827_{-0.0199}^{-0.0207}$ & $0.6485$ & $0.7154$ \\
$Y_\mathrm{He}$ &$0.248$ & $0.2479_{-7.1e-05}^{+7.3e-05}$ & $0.2478$ & $0.2481$ \\
$H_0$ &$67.428$ & $67.763_{-1.578}^{+1.152}$ & $65.246$ & $70.426$ \\ 
$10^{+9}A_\mathrm{s }$ &$1.987$ & $1.993_{-0.033}^{+0.03}$ & $1.931$ & $2.06$ \\
$\sigma_8$ &$0.7913$ & $0.797_{-0.0228}^{+0.0242}$ & $0.7739$ & $0.8416$ \\
$S_8$ &$0.819$ & $0.819\pm0.012$ & $0.795$ & $0.844$ \\
$A_\mathrm{planck}$ &$1.00158$ & $1.00196_{-0.00245}^{+0.00226}$ & $0.99777$ & $1.00649$ \\
\hline
\end{tabular}
\end{table*}
\raggedbottom

\pagebreak

\clearpage

\bibliography{apssamp}

\end{document}